\documentclass[prd,twocolumn,nofootinbib]{revtex4}
\usepackage{graphicx, epsfig}
\usepackage{color}
\usepackage{mathrsfs}
\usepackage{bm}
\usepackage{amsmath,amssymb}
\usepackage{mathrsfs}
\usepackage[caption=false]{subfig}
\usepackage{hyperref}
\usepackage[normalem]{ulem}
\usepackage{mathtools}
\usepackage[x11names]{xcolor}

\definecolor{fashionfuchsia}{rgb}{0.96, 0.0, 0.63}
\colorlet{no_so_fashion_purple}{blue!50!red}

\newcommand{\be}{\begin{equation}}
\newcommand{\ee}{\end{equation}}
\newcommand{\ba}{\begin{eqnarray}}
\newcommand{\ea}{\end{eqnarray}}

\newcommand{\nn}{\nonumber}
\newcommand{\half}{\frac{1}{2}}
\newcommand{\fourth}{\frac{1}{4}}

\newcommand{\hatn}{\hat n}
\newcommand{\calA}{\mathscr{A}}
\newcommand{\calD}{\mathscr {D}}
\newcommand{\Lag}{{\cal L}}
\newcommand{\Ham}{{\cal H}}

\newcommand{\bfW}{{\bf W}}

\def\half{\frac{1}{2}}

\begin{document}
\title{Unexciting non-Abelian electric fields}
\author{Tanmay Vachaspati}
\affiliation{
$^*$Physics Department, Arizona State University, Tempe,  Arizona 85287, USA.
}

\begin{abstract}
Electric fields in QED are known to discharge due to Schwinger pair production of
charged particles. Corresponding electric fields in non-Abelian theory are known to discharge 
due to the production of gluons.  Yet electric flux tubes in QCD ought to be stable to the
production of charged gluons as they confine quarks. We resolve this conundrum by
finding electric field configurations in pure non-Abelian gauge theory in which the 
Schwinger process is absent and the electric field is protected against quantum dissipation.
We comment on the implications for QCD flux tubes.
\end{abstract}

\maketitle

\section{Introduction}
\label{introduction}

Quantum particle production in time-dependent backgrounds continues to be a 
topic of great interest. The situation arises in the context of gravitational collapse
and leads to Hawking radiation~\cite{Hawking:1975vcx}, in cosmology where particle creation occurs due
to the expansion of the universe~\cite{Birrell:1982ix}, and in Schwinger pair production~\cite{Schwinger:1951nm} 
when the electric field is described in terms of a time-dependent gauge field. The 
production of particles implies that there is backreaction on the background, and 
external agencies must maintain the background or else it will dissipate. For example, 
black holes evaporate and capacitors discharge. 

In a recent paper~\cite{Vachaspati:2022ayz} 
we discussed time-dependent backgrounds that are ``unexciting'', 
{\it i.e.} time-dependent backgrounds in which there is no net production of particles.
(This is connected to ``shortcuts to adiabaticity (STA)'' in quantum mechanical systems
reviewed in~\cite{RevModPhys.91.045001}, and also related to
certain gravitational systems discussed in~\cite{Parikh:2011aa,Parikh:2012ny}.)
In most such backgrounds, particles are produced and then later absorbed so that the 
net particle production vanishes.
A subset of unexciting backgrounds are those for which the particle production
vanishes at all times. Such backgrounds are of interest because they are protected 
against quantum dissipation and no external agency is required to maintain the
background. Their time-dependence is of a stationary nature. An
example is that of a boosted soliton that is coupled to other quantum degrees of 
freedom: a boosted soliton is time-dependent but does not radiate particles. 

Here we are interested in 
electric field backgrounds in pure non-Abelian gauge theory. Generally we
would expect such electric field backgrounds to discharge due to the
Schwinger pair production of gluon excitations. However, confinement
suggests that electric flux tube configurations should be protected
against quantum dissipation. 
By carefully choosing the time-dependence of the
electric field it is possible to suppress particle production in 
a given excitation mode~\cite{Kim:2011jw,Vachaspati:2022ayz}, yet
it is unclear what, if
anything, could prevent Schwinger pair production completely. Even an exponentially 
suppressed pair production rate would eventually cause the electric flux tube to
dissipate.

To clarify this motivation further, consider Schwinger pair production in the case
of electrodynamics with a uniform electric field of strength $E$ and when the charge 
carriers have mass $m_e$ and charge $e$. The rate of particle production goes 
as~\cite{Schwinger:1951nm}, 
\be
{\dot n} \propto e^2 E^2 \exp(-\pi m_e^2/eE)
\label{schwinger}
\ee
and can be understood in different ways depending on the choice of gauge.

\begin{figure}
\includegraphics[width=0.45\textwidth,angle=0]{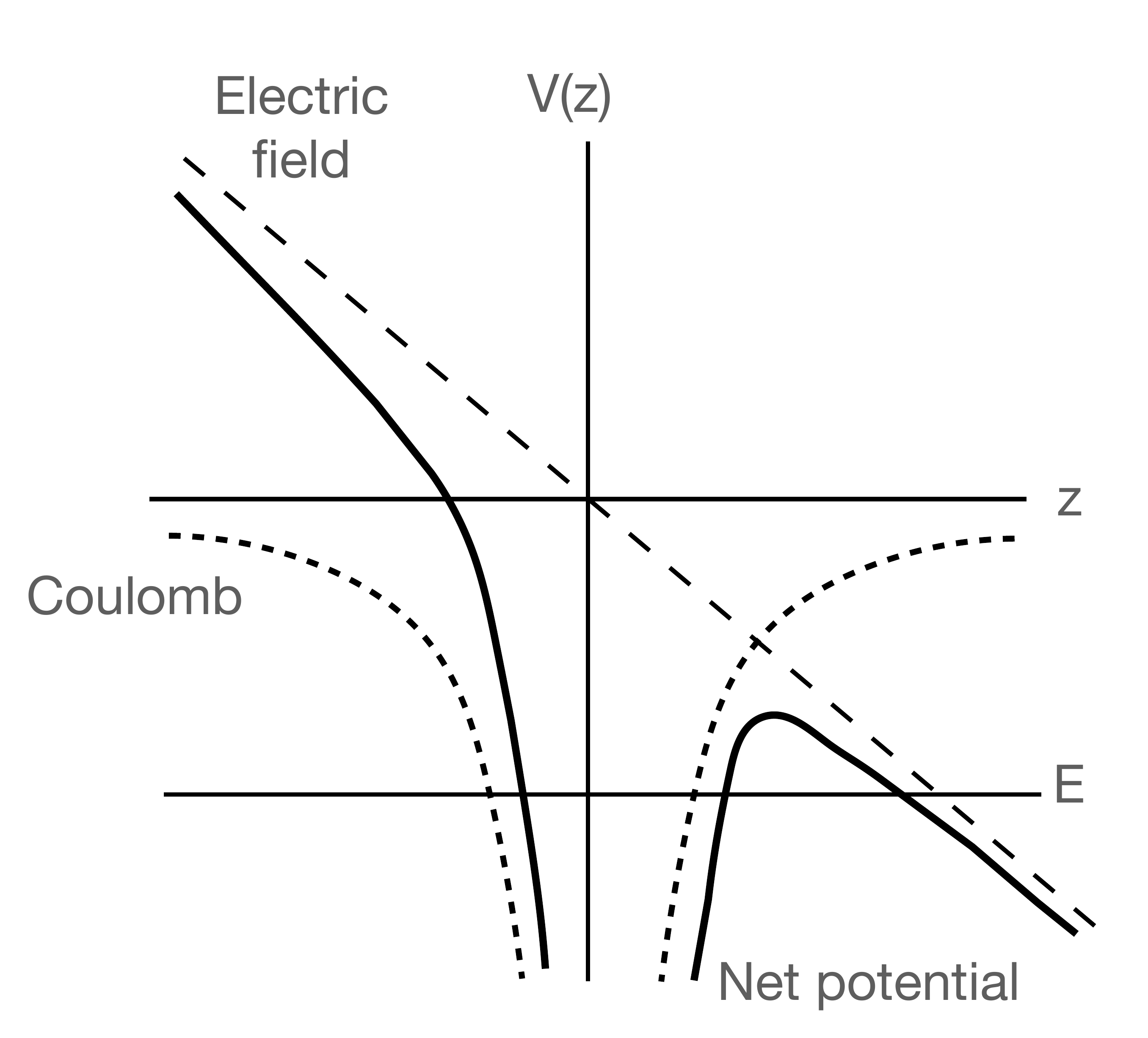}
 \caption{A sketch for the tunneling picture of Schwinger pair production.
 The Coulomb curves (small-dashed) show the Coulomb attractive potential
 between the pair of charges; the long dashed line shows the potential 
 energy of the positive charge
 due to the external electric field, and the solid curves are the sum of the 
 Coulomb and external electric potentials. Quantum fluctuations deep 
 in the potential well with energy E can tunnel out and escape to infinity
 and would be interpreted as Schwinger particle pairs.}
\label{tunnel}
\end{figure}

If one adopts Coulomb gauge, the gauge potential for a uniform electric field along
the $z-$direction is
\be
A^\mu = (-E z,0,0,0)
\ee
Then the gauge potential diverges asymptotically in the $z$ direction. As discussed
in Ref.~\cite{Itzykson:1980rh,Kim:2000un,Kim:2003qp}
for example, particle production can be viewed as
a tunneling process in a potential which is infinitely negative as $z\to \infty$
(see Fig.~\ref{tunnel}) so that the produced particles escape to infinity.
For an electric field in a finite but large domain, the potential does not
diverge but goes to a constant as $z \to \infty$ and the particles still escape
to infinity.

Alternatively, if one adopts temporal gauge, as we shall do, the gauge potential is
\be
A^\mu = (0,0,0,E t)
\ee
Now the gauge field is spatially well-behaved but varies with time.
Any quantum excitations of charged fields will obtain time-dependent
frequencies, just as for a simple harmonic oscillator with a time-varying
spring constant. The charged quantum modes in the vacuum will get excited 
due to this time-dependent background leading to pair production. The rate of 
particle production can
be calculated using the standard machinery of Bogolyubov 
coefficients ({\it e.g.}~\cite{Bogolyubov:1958,Birrell:1982ix}), or in the framework of the 
``classical-quantum correspondence'' where quantum particle production 
is described in terms of solutions of the classical 
equations~\cite{Vachaspati:2018llo,Vachaspati:2018hcu}.

Here we consider a pure non-Abelian SU(2) gauge theory with a background (``color'')
electric field in temporal gauge. In addition, the theory contains ``gluon'' excitations
that are massless and charged. A background electric field that is analogous to that
in ordinary electrodynamics is known to pair produce 
gluons~\cite{Matinyan:1976mp,Brown:1979bv,Yildiz:1979vv,Ambjorn:1981qc,
Ambjorn:1982nd,Nayak:2005yv,Cooper:2005rk,Cooper:2008vy,Nair:2010ea,
Kim:2011jw,Ilderton:2021zej,Huet:2014mta,Ragsdale:2017wgi,Karabali:2019ucc,
Cardona:2021ovn}.
One important difference from the original Schwinger calculation is that the gluons are
massless and the exponential suppression in \eqref{schwinger} is absent. In fact,
there are ultraviolet and infrared divergences as discussed in Ref.~\cite{Cardona:2021ovn}
that are presumably controlled by asymptotic freedom and confinement. However,
it appears that no matter how weak the electric field strength is, there is always
some particle production and hence the electric field should decay. 

If any non-Abelian electric field decays due to the Schwinger process, it would
imply that any external electric charge would get shielded by gluons and the resulting
long range electric field would vanish. This runs counter to the picture that QCD has
electric flux tubes that confine electric charges, and we are led to the question if 
there can be non-Abelian electric field configurations that are immune to the
Schwinger process, {\it i.e.} non-Abelian electric fields that are unexciting.
Such electric flux tubes would be models for the QCD string responsible for
confinement that have been discussed now 
for nearly half a 
century~\cite{Kogut:1974sn,Kogut:1974ag,Takahashi:2002bw,Bissey:2006bz}.

A guess for an unexciting non-Abelian electric field configuration was suggested
in Ref.~\cite{Vachaspati:2022ayz}. One needs the electric field background to be stationary. 
Already we have mentioned boosted solitons as unexciting backgrounds. An 
alternative is to have ``rotating'' backgrounds. In non-Abelian gauge theories,
for example when quantizing magnetic monopole backgrounds, it is known that
there are rotor degrees of freedom that, when excited, endow a monopole with 
electric charge and convert it into a dyon. Could such rotor degrees of freedom
be relevant for unexciting non-Abelian electric fields?

Approaching the problem from a different point of view, one wishes to construct
``stationary'' gauge fields that lead to a uniform electric field. Fortunately this problem has
been analyzed in detail in Ref.~\cite{Brown:1979bv}
and it is found that there are two gauge inequivalent classes of
gauge fields that lead to
the same non-Abelian electric field. One of these ways is analogous to the
Abelian gauge potential, while the second one is necessarily due to the 
non-Abelian nature of the model. We will explain this in more detail in
Sec.~\ref{electricfield} but suffice it to say that this second description of
the electric field corresponds precisely to the uniform rotation of a rotor
degree of freedom with quantized angular momentum (Sec.~\ref{symmetry}). 
The analysis of Sec.~\ref{expansion} shows
explicitly that this gauge background is stationary and does not lead to
particle production, and consequently is protected against quantum dissipation.
Quantum excitations on top of the classical background will settle into some ground 
state which is very difficult to determine because of the strongly 
coupled nature of the system but, whatever the state may be, it will be stationary.
In Sec.~\ref{k0modes} we discuss the
simpler quantization of the homogeneous modes in
the linearized approximation. Even this limited analysis has some novel features. 
Most of our analysis is done for a uniform electric field as this is simpler but in 
Sec.~\ref{profile} we remark on strategies to determine the profile of a flux tube.
We start our discussion with a motivating illustration of an unexciting electric field 
in 1+1D in Sec.~\ref{1+1D} and summarize our conclusions in Sec.~\ref{conclusions}.

\section{An illustrative example in 1+1D}
\label{1+1D}

An example of an electric field configuration without Schwinger pair production
is already known in massless QED in 1+1 dimensions~\cite{Chu:2010xc,Gold:2020qzr}
with action,
\be
S= \int d^2x \left [ {\bar \psi} \gamma^\mu (i \partial_\mu + e A_\mu) \psi - 
\frac{1}{4} F_{\mu\nu}F^{\mu\nu} \right ]
\label{1+1action}
\ee
where $\psi$ is a fermion field, $A_\mu$ is a U(1) gauge field, and 
$F_{\mu\nu}$ is the field strength.

An unexciting electric field background is given by~\cite{Chu:2010xc},
\ba
F_{01} &=& Q ( \Theta (x+L/2) - \Theta (x-L/2) ) \nonumber \\
&& + g( f (x+L/2) - f(x-L/2) )
\label{efield}
\ea
where 
$g=e/\sqrt{\pi}$ and $Q$ is the external charge
on a capacitor with plate separation $L$, and
\be
f(x) = -\frac{Q}{2g} {\rm sgn}(x) \left ( 1- e^{-g |x|} \right ).
\ee
The terms proportional to $Q$ in \eqref{efield} give the electric field of the
classical capacitor, while the last two terms give the contribution of a quantum
condensate of fermions. 
Quantum effects provide extra sources that screen some of the classical electric field, 
resulting in a net electric field in which there is no Schwinger pair production.

Similarly in the non-Abelian case discussed below, we consider an electric field
configuration that solves the classical equations of motion only in the presence
of some sources (see Sec.~\ref{expansion}). These sources can be external or 
be generated internally by quantum effects due to higher order interactions.

\section{Uniform electric field}
\label{electricfield}

As discussed in~\cite{Brown:1979bv}, a homogeneous non-Abelian electric field
can be derived from several gauge inequivalent potentials. Say we want the gauge
potentials for an electric
field in the third isospin direction and pointing along the $z$ direction, {\it i.e.}
$E \equiv E^3_z = W^3_{0z}$ where $W^a_\mu$ are gauge potentials from which
the field strength $W^a_{\mu\nu}$ is derived in the usual way,
\be
W^a_{\mu\nu} = \partial_\mu W^a_\nu - \partial_\nu W^a_\mu +
\epsilon^{abc} W^b_\mu W^c_\nu
\ee
where we have set the gauge coupling to unity since we will only be considering 
non-interacting quantum fluctuations on an electric field background.

The first ``trivial'' way to obtain $E^3_z$ is to take,
\be
W^a_\mu = - E\, t \delta^{a3} \partial_\mu z
\label{W1}
\ee
Then
\be
E^a_i \equiv -(\partial_t W^a_i - \partial_i W^a_t + \epsilon^{abc} W^b_t W^c_i )
= E \delta^{a3} \delta_{iz}
\ee

A second way to obtain $E^3_z$ is to take~\cite{Brown:1979bv},
\be
W^1_\mu = (\Omega, 0,0,0), \ \ 
W^2_\mu = (0,0,0, -E/\Omega), \ \ 
W^3_\mu =0
\label{W2}
\ee
where $\Omega$ is a constant 
(that will turn out to be quantized in Sec.~\ref{symmetry}.)
With \eqref{W2} we also obtain
\be
E^a_i = E \delta^{a3} \delta_{iz} .
\label{Etoo}
\ee
Even though the gauge potentials in Eqs.~\eqref{W1} and \eqref{W2} yield the same
field strength, they are gauge inequivalent, as are the gauge fields for different values
of $\Omega$, as can be seen by computing other gauge invariant quantities 
such as~\cite{Brown:1979bv}\footnote{We are using the signature $(+,-,-,-)$.},
\be
(D_\nu W^{\mu\nu})^a (D_\sigma W^{\ \sigma}_{\mu})^a 
= - E^2 \left ( \Omega^2 - \frac{E^2}{\Omega^2} \right ) 
\label{dwdw}
\ee
where
\be
(D_\nu W^{\mu\nu})^a \equiv \partial_\nu W^{\mu\nu a} + \epsilon^{abc} W_\nu^b W^{\mu\nu c}.
\label{covderiv}
\ee
Since the gauge invariant quantity on the left-hand side of \eqref{dwdw} depends
on $\Omega$, gauge fields for different $\Omega$ values are gauge inequivalent.
However, the energy density of the configuration is independent of $\Omega$, since 
the electric field does not depend on $\Omega$.

In the context of ``unexciting'' backgrounds discussed in~\cite{Vachaspati:2022ayz}, 
we would like to work in temporal gauge and set $W^a_0=0$. A gauge transformation
yields,
\be
\bfW'_\mu = U^{-1} \bfW_\mu U + i U^{-1} \partial_\mu U
\ee
where now $\bfW_\mu \equiv W^a_\mu T^a = W^a_\mu \sigma^a/2$, $T^a$ are
the generators of SU(2) normalized to ${\rm Tr}(T^aT^b)=\delta^{ab}/2$ and
$\sigma^a$ are the Pauli spin matrices.
The gauge transformation $U$ that takes us to temporal gauge is given by,
\be
U = e^{i \sigma^1 \Omega t /2}
\ee
Using the identities,
\ba
e^{i \hatn \cdot \sigma \theta} &=& \cos\theta + i \hatn\cdot\sigma \sin\theta \nn \\
e^{i \hatn \cdot {\vec \sigma} \theta} {\vec \sigma} e^{-i \hatn \cdot {\vec \sigma} \theta} &=&
{\vec \sigma} \cos(2\theta ) + {\hat n} \times {\vec \sigma} \sin(2\theta) 
\nn \\ 
&& 
+ {\hat n}\, {\hat n}\cdot {\vec \sigma} ( 1-\cos(2\theta) )
\ea
we find
\ba
W^1_\mu{}' &=& 0 \nn \\ 
W^2_\mu{}' &=& -\frac{E}{\Omega} \cos (\Omega t) \partial_\mu z , \nn \\ 
W^3_\mu{}' &=& -\frac{E}{\Omega}  \sin (\Omega t) \partial_\mu z .
\ea
A global SU(2) rotation can be used to bring the gauge fields to a form that we will use
\ba
W^1_\mu &=& -\frac{E}{\Omega} \cos (\Omega t) \partial_\mu z  , \nn \\ 
W^2_\mu &=& -\frac{E}{\Omega}  \sin (\Omega t) \partial_\mu z , \nn \\
W^3_\mu &=& 0
\label{bkgnd}
\ea
and we have dropped the primes for convenience. 

Ref.~\cite{Brown:1979bv} wrote the gauge field in the form of \eqref{W2} to show
that the same electric field can be obtained with a one parameter ($\Omega$) family
of gauge inequivalent gauge fields. We are only interested in a fixed value of
$\Omega$ and it is more convenient to write the non-vanishing components as
\be
W^1_\mu = -\epsilon \cos (\Omega t) \partial_\mu z  , \
W^2_\mu = -\epsilon  \sin (\Omega t) \partial_\mu z .
\label{bkgnd2}
\ee
Then the electric field is given by 
\be
E=\Omega \epsilon .
\label{Ealphaeps}
\ee
Equivalently, we can work
with $W^\pm_\mu \equiv W^1_\mu \pm i W^2_\mu$,
\be
W^\pm_\mu = -\epsilon  e^{\pm i \Omega t} \partial_\mu z , \ \ 
W^3_\mu =0.
\label{bkgndpmW}
\ee
Defining
\be
W_{\mu\nu}^\pm = W_{\mu\nu}^1 \pm i W_{\mu\nu}^2
\ee
we find
\be
W_{\mu\nu}^\pm = \partial_\mu W_\nu^\pm - \partial_\nu W_\mu^\pm 
\pm i (W_\mu^3 W_\nu^\pm - W_\nu^3  W_\mu^\pm )
\label{Wmnpm}
\ee
and
\be
W_{\mu\nu}^3 = \partial_\mu W_\nu^3 - \partial_\nu W_\mu^3 
+ \frac{i}{2} (W_\mu^+ W_\nu^- - W_\nu^+ W_\mu^- ) .
\label{Wmn3}
\ee
For the background in \eqref{bkgnd} we get,
\be
W_{\mu\nu}^\pm = \mp i E (\partial_\mu t \, \partial_\nu z 
                                            - \partial_\nu t \, \partial_\mu z ) e^{\pm i\Omega t}, \ \ 
W_{\mu\nu}^3 = 0
\label{Wmunua}
\ee

The Lagrangian density for the non-Abelian gauge sector is
\be
\Lag_g = - \frac{1}{4}  \left [ W_{\mu\nu}^+ W^{\mu\nu -} + W_{\mu\nu}^3 W^{\mu\nu 3} \right ]
\label{lag}
\ee
The full model will necessarily include a Lagrangian density for external sources and their
couplings to the gauge fields as we will discuss in Sec.~\ref{expansion}.

The important lesson of this section is that in non-Abelian gauge theories we have
several gauge inequivalent choices for the gauge potential corresponding to an electric 
field background.
The gauge potential of interest to us is time-dependent only because it is rotating in 
gauge field space as in \eqref{bkgndpmW}.

\section{Expansion around a flux tube background}
\label{expansion}

We now consider the electric field configuration in \eqref{bkgnd} 
as a background and denote it by $A_\mu$. We assume it is produced by some unspecified 
external sources, and we wish to determine if it leads to Schwinger pair creation.
We write
\be
W_\mu^\pm = A_\mu^\pm + e^{\pm i\Omega t}Q_\mu^\pm, \ \ W_\mu^3 = A_\mu^3 + Q_\mu^3
\label{WAQ}
\ee
where now $A_\mu^a$ includes an unspecified radial profile function, $f(r)$,
($r=\sqrt{x^2+y^2}$),
\be
A^\pm_\mu = -\epsilon  e^{\pm i \Omega t} f(r) \, \partial_\mu z , \ \ 
A^3_\mu =0,
\label{bkgndpm}
\ee
and $Q_\mu^a$
are quantum excitations around the background. More explicitly,
\be
W_\mu^\pm = e^{\pm i\Omega t} \left ( - \epsilon f(r)\, \partial_\mu z + Q_\mu^\pm \right ), \ \ 
W_\mu^3 = Q_\mu^3
\ee
Then
\ba
W_{\mu\nu}^\pm &=& e^{\pm i\Omega t} \bigl [ {\calA}_{\mu\nu}^\pm +
({\calD}_\mu Q_\nu^\pm - {\calD}_\nu Q_\mu^\pm ) \nn \\
&& \hskip 2 cm
\pm i (Q_\mu^3 Q_\nu^\pm - Q_\nu^3 Q_\mu^\pm ) \bigr ]
\ea
where 
\ba
{\calA}_{\mu\nu}^\pm &\equiv& \pm i E f(r) ( \partial_\mu z \partial_\nu t  - \partial_\nu z \partial_\mu t )
\nn \\
&& \hskip 1.5 cm
+ \epsilon f'(r) ( \partial_\mu z \partial_\nu r  - \partial_\nu z \partial_\mu r ),
\label{Amunu}
\ea
\be
{\calD}_\mu Q_\nu^\pm \equiv \partial_\mu Q_\nu^\pm 
\pm i \Omega \partial_\mu t \, Q_\nu^\pm
\pm i \epsilon f(r) \partial_\mu z \, Q_\nu^3 .
\label{calDQ}
\ee
We also have
\be
W_{\mu\nu}^3 = {\calD}_\mu Q_\nu^3 - {\calD}_\nu Q_\mu^3 
+ \frac{i}{2} (Q_\mu^+ Q_\nu^- - Q_\nu^+ Q_\mu^- )
\ee
with
\ba
{\calD}_\mu Q_\nu^3 &\equiv& \partial_\mu Q_\nu^3 
+ i\frac{\epsilon}{2} f(r) \partial_\mu z \, Q_\nu^+ 
- i\frac{\epsilon}{2} f(r)  \partial_\mu z \, Q_\nu^- \nn \\
&=& \partial_\mu Q_\nu^3 - \epsilon f(r) \partial_\mu z Q_\nu^2
\label{calDQ3corrected}
\ea

The Lagrangian density for $Q^a_\mu$ is evaluated from \eqref{lag},
\ba
\Lag_g &=& - \frac{1}{4} \biggl | {\calA}_{\mu\nu}^+ +
({\calD}_\mu Q_\nu^+ - {\calD}_\nu Q_\mu^+ ) \nn \\
&& \hskip 2.5 cm
+ i (Q_\mu^3 Q_\nu^+ - Q_\nu^3 Q_\mu^+ ) \biggr |^2 \nn \\
&& \hskip -0.5 cm
- \frac{1}{4} \biggl | {\calD}_\mu Q_\nu^3 - {\calD}_\nu Q_\mu^3 
+ \frac{i}{2} (Q_\mu^+ Q_\nu^- - Q_\nu^+ Q_\mu^- ) \biggr |^2
\label{LAQ}
\ea

The Lagrangian density \eqref{LAQ} describes the interaction of $Q_\mu^a$
with the background $A_\mu$. 
Even at this stage it is clear that there can be no particle production: 
the background dependent terms in \eqref{LAQ}, for example
${\calA}_{\mu\nu}^\pm$ and in ${\calD}_\mu Q_\nu^a$, 
are all independent of time
and there is no $z-$dependence either.
For Schwinger pair production, the gauge field background should either have 
non-trivial time-dependence or it should have non-trivial spatial dependence, 
as discussed in Sec.~\ref{introduction}. 
The interactions of $Q_\mu^a$ with the background will lead to a non-trivial
ground state wavefunctional but without any time-dependent excitations.
Hence the electric-magnetic field background 
in \eqref{Amunu} is ``unexciting''. 

We will now examine the system more explicitly by expanding the Lagrangian density 
in powers of $Q$.

\subsection{First order Lagrangian density}
\label{firstorder}

The Lagrangian density up to linear order terms in $Q$ is,
\ba
\Lag_g^{(1)} &=&
- \frac{1}{4} \calA_{\mu\nu}^+ \bigl ( {\calD}^\mu Q^{\nu +} - {\calD}^\nu Q^{\mu +}   \bigr )^*
+ {\rm c.c.} \nn \\
&\to& 
+ \frac{1}{2} ( {\calD}^\mu \calA_{\mu\nu}^+ ) Q^{\nu - }
+ {\rm c.c.}
\label{LAQ1}
\ea
where in the second line we have dropped total derivative terms.
The linear order variation doesn't vanish; neither do they for the illustrative
example in \eqref{efield} since there are external sources for the electric field
and quantum effects induce a condensate of sources. 
Here we simply supplement this linear order Lagrangian density with a source term 
that couples external currents, $j_\mu^a$, to $W_\mu^a$,
\be
\Lag_s = j_\mu^a W^{\mu a} = j_\mu^a A^{\mu a} + e^{\pm i\Omega t}  j_\mu^\mp Q^{\mu\pm}
+j_\mu^3 Q^{\mu 3}
\ee
The external current includes sources that are necessary to generate the 
background net flux of electric field. The currents can also contain
effective quantum contributions that arise due to higher order interactions as
these backreact at the linear level (see Sec.~\ref{1+1D}).
Here we simply assume the existence of such a current without any dynamical 
explanation. 

Requiring that the linear order terms vanish up to total derivatives gives, 
\ba
j_\mu^\pm &=& -\frac{\epsilon}{2} e^{\pm i\Omega t}
                      \left ( f'' + \frac{f'}{r} + \Omega^2 f \right ) \partial_\mu z, \nn \\ 
j_\mu^3 &=& -\Omega \epsilon^2  f^2 \, \partial_\mu t.
\label{jmu}
\ea
These currents do not include the external sources located at $z = \pm \infty$.
To see this it is most transparent to consider the Abelian case of a homogeneous
electric field with field strength tensor
\be
F_{\mu\nu} = E ( \partial_\mu t \partial_\nu z - \partial_\nu t \partial_\mu z ).
\ee
Insertion into Maxwell's equations gives 
\be
j_\mu = -\partial^\nu F_{\nu\mu} = 0
\ee
and the charged capacitor plates at $z=\pm \infty$ are not included in $j_\mu$.

Since the currents in \eqref{jmu} do not include the asymptotic sources, they must
arise entirely as a quantum condensate similar to the second line in \eqref{efield}.
To show that such a condensate arises due to the strong interactions is difficult
but some progress may be made in the semiclassical approximation. We start
with the equation of motion for the gauge fields,
\be
D_\nu W^{\nu\mu a}=0
\label{opeq}
\ee
where the covariant derivative is defined in \eqref{covderiv}. We
then substitute
\be
W_\mu^a = A_\mu^a + q_\mu^a
\ee
in \eqref{opeq} where $A^{\mu a}$ is classical 
while $q^{\mu a}$ is quantum. The operator equation \eqref{opeq} is replaced
by its ground state expectation value. If we assume that expectation values
of odd powers of $q^{\mu a}$ vanish, some algebra yields,
\be
D_\nu^{(A)} A^{\nu\mu a} = -j^{\mu a}
\ee
where $D_\nu^{(A)} A^{\nu\mu a}$ is as in \eqref{covderiv} with $W$ replaced by $A$ and
\ba
j^{\mu a} &\equiv& \epsilon^{abc} \langle \partial_\nu q^{\nu b} q^{\mu c}  
- q^{\nu b} \partial^\mu q_\nu^c + 2 q^{\nu b} \partial_\nu q^{\mu c} \rangle_R \nn \\
&& 
+ A_\nu^b \langle q^{\nu a}q^{\mu b} - 2 q^{\nu b} q^{\mu a} \rangle_R
+ A_\nu^a \langle q^{\nu b} q^{\mu b} \rangle_R \nn \\
&&
+ A^{\mu b} \langle q^b_\nu q^{\nu a} \rangle_R - A^{\mu a} \langle q_\nu^b q^{\nu b} \rangle_R
\label{jmua}
\ea
The symbol $\langle \cdot \rangle_R$ refers to the renormalized vacuum
expectation value. One possible renormalization scheme would be to 
subtract out the expectation value in the trivial vacuum with zero electric field,
\be
\langle {\cal O} \rangle_R = \langle {\cal O} \rangle - \langle {\cal O} \rangle_{\epsilon=0}
\label{renorm}
\ee
for any quantum operator ${\cal O}$.
To illustrate how the current might
result in a charge condensate, consider the $\mu=0$, $a=3$ component of \eqref{jmua}.
Since we are working in temporal gauge and $q_\mu^a$ is defined in
terms of $Q_\mu^a$ using \eqref{WAQ}, the expression simplifies,
and we get
\ba
j^{(3)}_0 &=& \Omega \langle (Q_i^{(1)})^2 + (Q_i^{(2)})^2 \rangle_R \nn \\
&&
+ \langle Q_i^{(1)} {\dot Q}_i^{(2)} - Q_i^{(2)} {\dot Q}_i^{(1)} \rangle_R
\ea
Assuming the ground state of $Q_\mu^a$ has zero ``angular
momentum'' (also see Sec.~\ref{k0modes}), we get a
charge density,
\ba
j^{(3)}_0 = \Omega \langle (Q_i^{(1)})^2 + (Q_i^{(2)})^2 \rangle_R .
\ea
This can be compared to \eqref{jmu} and for a homogeneous electric field
($f=1$) we have,
\be
\epsilon^2 = - \langle (Q_i^{(1)})^2 + (Q_i^{(2)})^2 \rangle_R .
\ee
Note that the right-hand side can be positive due to the definition in \eqref{renorm}.

In summary, the background need not satisfy the classical equations of motion;
instead it is more realistic to ask if the background satisfies the semiclassical
equations of motion. By examining the semiclassical equations, we discover
that quantum effects can indeed provide appropriate sources for the electric
field background under consideration.

\subsection{Second order Lagrangian density}
\label{secondorder}

The quadratic order Lagrangian density is,
\ba
\Lag_g^{(2)} &=&
- \frac{1}{4}
\bigl | {\calD}_\mu Q_\nu^+ - {\calD}_\nu Q_\mu^+   \bigr |^2 
- \frac{1}{4} \bigl | {\calD}_\mu Q_\nu^3 - {\calD}_\nu Q_\mu^3  \bigr |^2 \nn \\
&&
+ \frac{i}{4} \left [ \calA_{\mu\nu}^+ (Q^{\mu 3} Q^{\nu -} - Q^{\nu 3} Q^{\mu -} ) - c.c. \right ] .
\label{LAQ2}
\ea
or, explicitly,
\ba
&& \hskip -0.75 cm 
\Lag_g^{(2)} = \nn \\
&& 
\half ( {\dot Q}^{(1)}_i - \Omega Q^{(2)}_i )^2
+\half ( {\dot Q}^{(2)}_i + \Omega Q^{(1)}_i )^2 
+\half ( {\dot Q}^{(3)}_i  )^2 \nn \\
&-& \fourth ( \partial_i Q^{(1)}_j -  \partial_j Q^{(1)}_i )^2 \nn \\
&-& \fourth ( \partial_i Q^{(2)}_j -  \partial_j Q^{(2)}_i + \epsilon f ({\hat z}_i Q^{(3)}_j -{\hat z}_j Q^{(3)}_i ))^2 \nn \\
&-& \fourth ( \partial_i Q^{(3)}_j -  \partial_j Q^{(3)}_i - \epsilon f ({\hat z}_i Q^{(2)}_j -{\hat z}_j Q^{(2)}_i ))^2 \nn \\
&+& \epsilon f' ({\hat z}_i {\hat r}_j - {\hat z}_j {\hat r}_i ) Q^{(2)}_i Q^{(3)}_j
\label{Lg2corrected}
\ea
where $Q^\pm_i = Q^{(1)}_i \pm i Q^{(2)}_i$, ${\hat z}_i$, ${\hat r}_i$ are unit vectors in the $z-$ and $r-$
directions, and the contraction of spatial indices is with the
Kronecker delta, {\it e.g.} $( {\dot Q}^{(3)}_i  )^2 =  {\dot Q}^{(3)}_i {\dot Q}^{(3)}_i $.

Denoting the momentum conjugate to $Q^a_i$ by $P^a_i$, the Hamiltonian density corresponding
to $\Lag_g^{(2)}$ is,
\ba
&& \hskip -0.75 cm 
\Ham_g^{(2)} = \half ( P^{(a)}_i )^2
+ \fourth ( \partial_i Q^{(1)}_j -  \partial_j Q^{(1)}_i )^2 \nn \\
&+& \fourth ( \partial_i Q^{(2)}_j -  \partial_j Q^{(2)}_i + \epsilon f ({\hat z}_i Q^{(3)}_j -{\hat z}_j Q^{(3)}_i ))^2 \nn \\
&+& \fourth ( \partial_i Q^{(3)}_j -  \partial_j Q^{(3)}_i - \epsilon f ({\hat z}_i Q^{(2)}_j -{\hat z}_j Q^{(2)}_i ))^2 \nn \\
&-& \epsilon f' ({\hat z}_i {\hat r}_j - {\hat z}_j {\hat r}_i ) Q^{(2)}_i Q^{(3)}_j + {\cal J}
\label{ham2}
\ea
where,
\be
{\cal J} = \Omega ( Q^{(2)}_i P^{(1)}_i - Q^{(1)}_i P^{(2)}_i ) 
\ee
is an ``angular momentum'' term.

To check that the background is unexciting, we simply note that the Hamiltonian density does
not have any time-dependence. If we decompose the field into eigenmodes, the
amplitude of each eigenmode is a quadratic variable that behaves like a simple harmonic 
oscillator variable with possibly an angular momentum contribution to the Hamiltonian
(see Sec.~\ref{k0modes}).

One potential complication is if the diagonalization of the second order
Hamiltonian density, $\Ham_g^{(2)}$, leads to modes that have imaginary frequencies,
{\it i.e.} are simple harmonic oscillators with inverted potentials. An analysis
of the spectrum of frequencies was carried out in 
Ref.~\cite{Bazak:2021xay} for the special choice $\Omega = \sqrt{E}$ 
and the authors found some unstable modes. These unstable modes imply that 
the quantum ground state of the $Q^a_i$ 
will be something other than the simple harmonic oscillator ground states
(at least for this choice of $\Omega$).
The instability will be tamed by the higher order interactions. Since the
higher order Lagrangian density is also time-independent (see Subsection.~\ref{L34} below),
there can be no particle production and the electric field background is
protected against quantum dissipation. 
However the danger is that the expectation value of the quantum excitations
might not vanish and then the separation between the classical background, $A_\mu^a$,
and the quantum excitations becomes unclear. The idea behind postulating $A_\mu^a$
as a background is that the quantum excitations are small and their expectation
values vanish. For this to happen, the background should not have any instabilities.
The stability analysis deserves to be investigated for the entire range of parameters.

To evaluate the quantum state of the 
$Q^a_i$ in the full interacting theory is beyond the scope of the present work,
but we discuss the simpler problem of the quantum state for homogeneous
modes in the quadratic Hamiltonian in Sec.~\ref{k0modes}. 

\subsection{Higher order Lagrangian density}
\label{L34}

For completeness we also give the cubic and quartic order terms in the Lagrangian 
density,
\ba
\Lag_g^{(3+4)} &=&
\frac{i}{4} \biggl [ 
({\calD}_i Q_j^+ - {\calD}_j Q_i^+ )
(Q_i^ 3 Q_j^- - Q_j^3 Q_i^- ) \nn \\
&& \hskip -1 cm
+ \frac{1}{2}  ( {\calD}_i Q_j^3 - {\calD}_j Q_i^3  )
(Q_i^+ Q_j^- - Q_j^+ Q_i^- ) - c.c. \biggr ] \nn \\
&& \hskip -1.75 cm
- \frac{1}{4} \bigl |Q_i^3 Q_j^+ - Q_j^3 Q_i^+  \bigr |^2
- \frac{1}{16} \bigl | Q_i^+ Q_j^- - Q_j^+ Q_i^- \bigr |^2
\label{LAQ3}
\ea

To summarize Sec.~\ref{expansion}, the electric field background \eqref{bkgndpm} is unexciting. 
There will be quantum fluctuations around this background and they will be in some stationary 
quantum ground state but particle production will be absent. For this reason, the electric
field background is stable to quantum dissipation.

\section{Symmetry and quantization}
\label{symmetry}

As discussed in the introduction, a static soliton can be boosted to give a
time-dependent background but this will not lead to particle production. 
This is because the static soliton has translational symmetry and a boosted
soliton provides time dependence to the background but the excitation
frequencies are still time-independent. In other words, the boosted soliton
is a ``stationary'' background.
Similarly, there should be a symmetry of the non-Abelian gauge theory,
and the electric field background we have been discussing should be due to 
a time-dependence in the symmetry variables. The situation is very similar
to the symmetries of monopoles and the rotor degree of freedom that 
leads to dyons, discussed for example in 
Refs.~\cite{Julia:1975ff,Callan:1982au} and extensively in 
Section 2.7 of~\cite{Preskill:1986kp}.

Consider the global gauge transformation
\be
W_\mu^\pm \to W'_\mu{}^\pm = e^{\pm i\theta} W_\mu^\pm, \ \ 
W_\mu^3 \to W'_\mu{}^3 = W_\mu^3
\ee
where $\theta$ is a constant parameter.
The boosted soliton background corresponds in this case to promoting the
rotor degree of freedom: $\theta \to \theta(t)$.

Next let
\be
W_\mu^\pm =  e^{\pm i\theta} v_\mu^\pm, \ \ 
W_\mu^3  = v_\mu^3
\label{thetav}
\ee
where $v_0^a=0$ in temporal gauge and we also assume that the $v_\mu^a$
background is static (just like the static soliton). The angular variable $\theta$
is assumed to only depend on time. Substituting \eqref{thetav} in the Lagrangian
for $W_\mu^a$ ({\it i.e.} \eqref{lag} integrated over space) gives the Lagrangian 
for $\theta$,
\be
L_\theta = \frac{1}{2} I {\dot \theta}^2
\label{Ltheta}
\ee
where the ``moment of inertia'' is
\be
I = \int d^3 x |v_i^+|^2 .
\ee
Solving for the quantum dynamics of $\theta$ is straightforward and leads to
the quantization of angular momentum and correspondingly
\be
\Omega = {\dot \theta} = \frac{l}{I}, \ \ l =0, \pm 1, \pm 2, \ldots
\label{alpha}
\ee
Thus the parameter $\Omega$ is quantized. The electric field strength is also 
quantized as
\be
E = \Omega \epsilon = \frac{\epsilon}{I} l.
\ee

The quantum number $l$ has the interpretation of the number of charge quanta.
To see this we evaluate the total charge, $q^a$, by integrating the $\mu=0$ component
of \eqref{jmu} over a volume $V$ for the homogeneous case ($f=1$). This gives
\be
q^a = - V \Omega \epsilon^2 \delta^{a3}= -l \delta^{a3}
\label{qa}
\ee
in units where the gauge coupling constant $g=1$ and we have used \eqref{alpha} and 
$I = \epsilon^2 V$.

\section{Flux tube profile}
\label{profile}

The quantum state of the fluctuations $Q^a_i$ will depend on the profile $f(r)$.
If we know the quantum state, we can calculate the expectation value of the Hamiltonian, 
$\langle H \rangle$, for the choice of $f(r)$. If $\langle H \rangle$ is minimized
for some $f(r)$, subject to a suitable constraint on the background electric field, 
it would specify the profile of the electric field flux tube and its tension.
We can define the background electric flux as
\be
\Phi^\pm \equiv \int d^2x\, {\calA}_{0z}^\pm = \mp i E e^{\pm i\Omega t} \int d^2x\, f(r),
\label{Phipm}
\ee
with ${\calA}_{\mu\nu}^\pm$ given in \eqref{Amunu} and $\Phi^3 =0$.
The flux itself is not gauge invariant but, having chosen a fixed form of the background
gauge fields, one can restrict the class of functions in this background gauge. 
This amounts to exploring profile functions
for which the integral on the right-hand side of \eqref{Phipm} is unity,
\be
2\pi \int dr\, r f(r) = 1.
\ee

Unfortunately, there is no simple way to find the quantum state of $Q^a_i$ for
given $f(r)$, especially as the fluctuations are strongly interacting. The only
hope seems to be to determine $f(r)$ numerically, using lattice techniques.

\section{Quantum state of the homogeneous modes}
\label{k0modes}

We now consider the Hamiltonian density in \eqref{ham2} for homogeneous excitations
on a uniform electric field background.
Then $\partial_i Q^a_j =0$ and $f=1$, and the Hamiltonian is related to the Hamiltonian
density by suitable factors of the spatial volume $V$,
\ba
&& 
\hskip -0.75 cm 
H_g^{(2)} = \frac{1}{2V} ( P^{(a)}_i )^2
+ \frac{\epsilon^2 V}{4} ({\hat z}_i Q^{(3)}_j -{\hat z}_j Q^{(3)}_i )^2 \nn \\
&& \hskip -0.85 cm 
+\frac{\epsilon^2 V}{4} ({\hat z}_i Q^{(2)}_j -{\hat z}_j Q^{(2)}_i )^2
+ \Omega ( Q^{(2)}_i P^{(1)}_i - Q^{(1)}_i P^{(2)}_i ) 
\label{ham2k0}
\ea
The Hamiltonian then can be written as a sum of six sub-Hamiltonians,
\ba
H_1 &=&  \frac{1}{2V} (P^{(1)}_x )^2 +  \frac{1}{2V} (P^{(2)}_x )^2 + \frac{\epsilon^2 V}{2}  ( Q^{(2)}_x )^2
\nn \\
&& \hskip 1 cm
+ \Omega ( Q^{(2)}_x P^{(1)}_x - Q^{(1)}_x P^{(2)}_x ) \\
H_2 &=&   \frac{1}{2V} (P^{(3)}_x )^2 + \frac{\epsilon^2 V}{2}  (Q^{(3)}_x )^2 \\
H_3 &=&  \frac{1}{2V} (P^{(1)}_y )^2 +  \frac{1}{2V} (P^{(2)}_y )^2 + \frac{\epsilon^2 V}{2}  ( Q^{(2)}_y )^2
\nn \\
&& \hskip 1 cm
+ \Omega ( Q^{(2)}_y P^{(1)}_y - Q^{(1)}_y P^{(2)}_y ) \\
H_4 &=&   \frac{1}{2V} (P^{(3)}_y )^2 + \frac{\epsilon^2 V}{2}  (Q^{(3)}_y )^2 \\
H_5 &=&  \frac{1}{2V} (P^{(1)}_z )^2 +  \frac{1}{2V} (P^{(2)}_z )^2 \nn \\
&& \hskip 1 cm
+ \Omega ( Q^{(2)}_z P^{(1)}_z - Q^{(1)}_z P^{(2)}_z ) \\
H_6 &=&  \frac{1}{2V} (P^{(3)}_z )^2
\ea
The Hamiltonians $H_2$ and $H_4$ correspond to simple harmonic
oscillators, while $H_6$ is that of a free particle. Their eigenstates are well-known;
the simple harmonic oscillators states are all bound, while the free particle only has
continuum states.
$H_5$ is the Hamiltonian for a free particle in two dimensions but with an additional 
angular momentum term, while $H_1$ and $H_3$ are identical in structure and
can also be thought of as a particle in two dimensions with an angular momentum
term plus an anisotropic potential.
Thus we only need to consider $H_1$ and $H_5$. 

Let us consider $H_5$ first and write it in a less cumbersome way,
\be
H_5 \to \half \pi_x^2 + \half \pi_y^2 + \Omega ( y \pi_x - x \pi_y )
\label{H5xy}
\ee
where $\pi_x$ and $\pi_y$ are momenta conjugate to $x$ and $y$, and we
have also chosen units so that $V=1$.
One can check that
\be
[ y \pi_x - x \pi_y , H_5]=0
\ee
and there are simultaneous eigenstates of the angular momentum operator,
$y \pi_x - x \pi_y$, and $H_5$. Defining polar coordinates in the usual way
\be
x = r\, \cos\theta, \ \ y = r\, \sin\theta
\ee
the energy eigenstates (up to an overall normalization factor) 
are given in terms of Bessel functions,
\be
\psi(r,\theta, t) = C
e^{-i{\cal E} t} J_l\left (\sqrt{2({\cal E}+\Omega m)} \, r \right ) e^{i m \theta},
\label{psirtheta}
\ee
with $m=0,\pm 1, \pm 2, \ldots$, $C$ is a normalization constant, and the energy, 
${\cal E}$, is constrained by ${\cal E} + \Omega m \ge 0$. 

An interesting point is that ${\cal E}$ can be negative. In fact, for $\Omega > 0$ and 
$m \to \infty$, ${\cal E}$ can be arbitrarily negative. We expect that when
the quartic interactions are taken into account, the energy may get bounded
from below. 
For example, if the interactions effectively provide a mass to the homogeneous
mode, we could consider a rotationally symmetric simple harmonic oscillator in 2D 
with an additional angular momentum term
\be
H_{\rm e.g.} = 
\half \pi_x^2 + \half \pi_y^2 + \frac{\mu^2}{2} (x^2+y^2) + \Omega ( y \pi_x - x \pi_y ),
\label{H5xyeg}
\ee
As the system has rotational symmetry, the angular momentum is still conserved.
The Schrodinger problem can be solved in polar coordinates by elementary means
to obtain the energy eigenvalues\footnote{We find it more convenient to not define the
principal quantum number to be $N=n+|m|$ as is conventionally done, in which case
$m=0,\pm 1, \ldots, \pm N$.},
\ba
E_{nm} &=& (n+|m|+1)\mu - \Omega m, \nn \\
&&
n=0,1,2, \ldots; \ \ m = 0, \pm 1, \pm 2, \ldots
\ea
Consider the case $\Omega > 0$. Then, for fixed $n$ and $m \to +\infty$, 
we have $E_{nm} \to (\mu - \Omega) m$,
and the energy is bounded from below only if $\mu \ge \Omega$. Considering that
$\Omega$ can have either sign, the
condition for the energy to be bounded from below is,
\be
\mu \ge |\Omega |
\label{condition}
\ee
and then the ground state is for $n=m=0$ with energy $\mu$.

In the notation of \eqref{H5xy}, $H_1$ can be written as
\be
H_1 \to \half \pi_x^2 + \half \pi_y^2 + \Omega ( y \pi_x - x \pi_y ) + \frac{\epsilon^2}{2} y^2
\label{H1xy}
\ee
The last term breaks rotational symmetry in the $xy-$plane and the angular
dependence will not be given by angular momentum eigenstates as in 
\eqref{psirtheta}. The wavefunction will now be peaked on the $x-$axis;
it will be bounded in the $y-$direction while remaining unbounded in the 
$x-$direction.

In principle, we can also treat the case of inhomogeneous excitations since
the Hamiltonian is quadratic. However, the diagonalization will be
technically difficult and the angular momentum term may resist complete
diagonalization as in the case of \eqref{H1xy}. 

In this section we have discussed quantum aspects of the homogeneous
quantum excitations. The energy spectrum for some of these homogeneous excitations 
is unbounded from below. If higher order interaction terms effectively give masses
to these excitation modes, the spectrum gets bounded provided the mass is greater
than $|\Omega|$ (see \eqref{condition}).

\section{Conclusions}
\label{conclusions}

We have found that an electric field in pure non-Abelian SU(2) gauge theory
can be stable against quantum dissipation 
to Schwinger pair production. (Similar constructions can be embedded in theories
with larger gauge groups.) 
This is because the gauge fields underlying the
electric field can be chosen as a stationary background in which a rotor degree
of freedom is rotating with fixed, quantized, angular momentum in internal space. The
quantum state of excitations on this stationary background is also stationary 
and there is no particle production and no dissipation. This resolves a conundrum 
that one might intuit from the case of Abelian electric fields, where the electric field
dissipates due to pair production, even if at an exponentially suppressed rate.
If the same conclusion applied to non-Abelian electric fields, QCD flux tubes
would be susceptible to decay due to Schwinger pair production of gluons.
Thus we conjecture that QCD flux tubes should be described by gauge fields
as given in Eq.~\eqref{bkgndpm}.
If we start with a non-Abelian electric field that is in the Abelian configuration 
of \eqref{W1},
a guess is that it would evolve into the unexciting non-Abelian configuration
of \eqref{bkgndpm}.

The existence of stable non-Abelian electric fields opens up a number of 
related questions. 
One issue we faced is that we had to postulate classical external charges to
source the background electric field. At present it is not known whether such 
sources need to be external or if they can arise due to the strong interactions
as discussed in Sec.~\ref{firstorder}. 
The key open question is to determine properties of the
ground state of quantum excitations around the background electric field.
In Sec.~\ref{secondorder} we have already pointed out the danger posed 
by unstable modes as then the separation between a classical background
and quantum fluctuations is not clear. Assuming that there is a
range of parameters where there are no unstable modes, there will be a
well-defined ground state
which will depend on the profile of the electric field flux tube.
Perhaps there are lattice techniques that can determine the optimum
flux tube profile for which the overall energy is minimized. The quantum state
can also settle the question whether the sources necessary for the 
stationary background can arise from the internal dynamics of
quantum excitations. These are fascinating but difficult questions that
we hope to examine in the future.

Another system of interest is that of dyons that carry non-Abelian electric
charge, as these can also arise as stationary solutions of non-Abelian Yang-Mills-Higgs
theories. We expect such dyons to be stable to Schwinger particle production 
of non-Abelian gauge bosons but a detailed analysis might yield surprises especially
for large $\Omega$, as in the discussion in Sec.~\ref{k0modes}.

\acknowledgements
I am grateful to Gia Dvali, Sang Pyo Kim, Parameswaran Nair, Igor Shovkovy,
Frank Wilczek and George Zahariade for helpful comments. T.V.  was supported 
by the U.S. Department of Energy, Office of High Energy 
Physics, under Award No.~DE-SC0019470.

\newpage

\bibstyle{aps}
\bibliography{paper}

\end{document}